\documentclass[10pt, conference]{IEEEtran}
\IEEEoverridecommandlockouts 
\IEEEpubid{\makebox[\columnwidth]{978-1-7281-8688-7/22/\$31.00~
\copyright{}2022 IEEE \hfill} \hspace{\columnsep}\makebox[\columnwidth]{}}

\usepackage{balance}
\usepackage{cite}
\usepackage{amsthm}
\usepackage{amsmath,amssymb,amsfonts}
\usepackage{algorithmic}
\usepackage{graphicx, subfigure}
\usepackage{textcomp}
\usepackage{xcolor}
\usepackage{balance}
\def\BibTeX{{\rm B\kern-.05em{\sc i\kern-.025em b}\kern-.08em
    T\kern-.1667em\lower.7ex\hbox{E}\kern-.125emX}}
    
\usepackage{cite}
\usepackage{amsmath,amssymb,amsfonts}
\usepackage{algorithmic}
\usepackage{graphicx}
\usepackage{textcomp}
\usepackage{tikz}

\usepackage{cite}
\usepackage{amssymb, amsmath}
\usepackage{lineno}
\usepackage{tikz}
\usepackage{dsfont}
\usepackage[ruled,vlined, linesnumbered]{algorithm2e}
\usepackage{math}
\usepackage{float}
\usepackage{times}
\usepackage{amssymb}
\usepackage{amsfonts}
\usepackage{caption}
\usepackage{pgfplots}
\usepackage{xcolor}
\usepackage{parskip}
\usepackage{cancel}
\usepackage[normalem]{ulem}
\usepackage[nodisplayskipstretch]{setspace}
\usepackage{framed}
\usepackage{lipsum}

\newcommand{\R}{\mathbb{R}}
\newcommand{\Prob}{\mathbb{P}}

\newcommand{\M}{\mathcal{M}}

\newcommand{\B}{\mathcal{B}}
\newcommand{\D}{\mathcal{D}}

\newcommand{\Prms}{\mathcal{P}}
\newcommand{\Map}{\mathbb{M}}

\newtheorem{theorem}{Theorem}

\definecolor{commentcol}{RGB}{205,133,63}

\begin{document}

\title{Agent-Based Simulations for Coverage Extensions in 5G Networks and Beyond}

\author{\IEEEauthorblockN{Chaima Ghribi}
\IEEEauthorblockA{\textit{Orange Labs} \\
Paris, France \\
Chaima.Ghribi@orange.com}
\and
\IEEEauthorblockN{Elie Cali}
\IEEEauthorblockA{\textit{Orange Labs} \\
Paris, France \\
Elie.Cali@orange.com}
\and
\IEEEauthorblockN{Christian Hirsch}
\IEEEauthorblockA{\textit{Bernoulli Institute} \\
\textit{University of Groningen}\\
Groningen, Netherlands\\
C.P.Hirsch@rug.nl}
\and
\IEEEauthorblockN{Benedikt Jahnel}
\IEEEauthorblockA{\textit{Weierstrass Institute for Applied }\\
\textit{Analysis and Stochastics}\\
Berlin, Germany \\
Benedikt.Jahnel@wias-berlin.de}
}

\maketitle

\begin{abstract}
Device-to-device (D2D) communications is one of the key emerging technologies for the fifth generation (5G) networks and beyond. It enables direct communication between mobile users and thereby extends coverage for devices lacking direct access to the cellular infrastructure and hence enhances network capacity.
D2D networks are complex, highly dynamic 
and will be strongly augmented by intelligence for decision making at both the edge and core of the network, which makes them particularly difficult to predict and analyze. Conventionally, D2D systems are evaluated, investigated and analyzed using analytical and probabilistic models (e.g., from stochastic geometry). However, applying classical simulation and analytical tools to such a complex system is often hard to track and inaccurate. 
In this paper, we present a modeling and simulation framework from the perspective of complex-systems science and exhibit an agent-based model for the simulation of D2D coverage extensions. 
We also present a theoretical study to benchmark our proposed approach for a basic scenario that is less complicated to model mathematically. Our simulation results show that we are indeed able to predict coverage extensions for multi-hop scenarios and quantify the effects of street-system characteristics and pedestrian mobility on the connection time of devices to the base station (BS). To our knowledge, this is the first study that applies agent-based simulations for coverage extensions in D2D.

\end{abstract}


\IEEEpeerreviewmaketitle

\section{Introduction}
\label{sec:intro}
\par The increasing complexity of mobile communication networks, due to pedestrian mobility, street systems, shadowing and interference, requires new modeling approaches and paradigms to predict and analyze them. 
Recently, mobile communication networks modeling and analysis has raised considerable interest, especially via applying complex-systems theory.

\par Agent-based modeling (ABM), cf.~\cite{Helleboogh,Olaf} is as a bottom-up modeling approach that considers a network of autonomous agents that interact with each other, and therefore represents an ideal framework to understand the interactions of heterogeneous nodes in a complex environment. In this paper, we investigate the ability of ABM to simulate device-to-device (D2D) communication networks in urban environments. We will mainly be interested here in the connectivity of the network modeled by a random graph, and we will not take into account the telecommunications phenomena (e.g., shadowing, interference, reflections, ...).

\par As a first step, we will see that ABM can completely reproduce and simulate a standard model of D2D networking, where the coverage extensions are provided by just one additional relaying hop. Further applying ABM to investigate coverage extension induced by D2D, we will extend the analytical result obtained for one hop, via simulations, to several hops. 

\par Then, as a second use case, we will show that ABM can simulate more complex environments, taking into account mobility of the users on real street systems, and taking advantage of the local ability of the system to let each agent take decisions locally depending on its environment. This will provide results on the influence of the type of street system considered on the mean communication time.

The paper is organized as follows. In Section~\ref{sec:relwork} we discuss related work, in Section~\ref{sec:model} we describe the ABM model, in Section~\ref{sec:simu} we describe the use cases investigated, in Section~\ref{sec:results} we comment on the results obtained, and in Section~\ref{sec:ccl} we give a short conclusion and some hints for future work.

\section{Related work}
\label{sec:relwork}
D2D multi-hop communication is often modeled and evaluated using stochastic geometry tools. Among the numerous subjects treated (cf.~for instance \cite{Ansari_survey}), mobility, coverage extension, and the influence of the street system are strongly underrepresented in the literature.

Regarding mobility, authors in \cite{Fedorov_SIR} have proposed an analytical approach to predict channel quality between two moving devices in a field of moving interfering stations. In \cite{Fedorov_mobility}, the authors present an approach for characterizing distributions of steady-state communications and outages as a function of various parameters using the uniform theory of kinetic equations. However, in both works, the analysis is focused on the impact of mobility on interference, and the support of mobility (namely street systems) is not taken into account.

Regarding coverage extensions, authors in \cite{Chakrabarti} propose a stochastic-geometry-based framework for evaluating the overall network coverage probability improvement due to minimization of congestion in the base station (BS) thanks to D2D.  
An energy-efficient two-hop D2D relay selection algorithm was proposed in \cite{Selmi_energy} to extend connectivity to the out-of-coverage devices. The proposed algorithm is distributed and based on the Stackelberg game in order to involve all the competing devices.
Authors in \cite{Babun_extendedcoverage} propose a system-level simulator to evaluate the performance of multi-hop communications via extensions of the coverage area of active macro BSs during emergency situations.
However, in all these works, the focus is put on techniques to improve the coverage extension. The mobility of users is rarely taken into account, and the urban environment is completely absent from the analysis.

Again, with the help of stochastic geometry, D2D connectivity has been studied for users placed on street systems (cf.~for instance \cite{LeGall}). However, authors there do not take into account mobility and coverage extensions.

Due to the complexity of D2D systems, stochastic geometry models will not be sufficient to capture the dynamics of the system and to model realistic scenarios while also considering urban environments and user mobility. Accurate models are very challenging to derive and their associated algorithms are complex to deploy. We believe ABM and simulation (cf.~for instance \cite{Klugl}), is a complementary and powerful tool that helps to address the trade-off between tractability and accuracy that affects stochastic-geometry models.

\section{ABM for coverage extensions in D2D}
\label{sec:model}

In this section, we give a detailed description of the D2D coverage-extension model in urban environments. In this ABM description, we consider two types of agents: BSs and devices. A base-station agent represents a fixed transceiver that serves as a central connection point for wireless mobile devices within its communication radius. Devices are represented as reactive agents that move in the environment and have a variety of different capabilities like neighborhood discovery and one-hop relay routing.  

In short terms, the system has the following composition. We consider a real urban environment of a street system. At initial time, BSs and devices are placed randomly on the streets, where we make the simplifying assumption that devices that are situated in buildings are not to be taken into account. This can be justified by the high frequencies used in 5G. The devices move independently and randomly at a constant speed. Moreover, two devices can communicate directly with each other if they are close enough. Let us note that this approach does not take shadowing nor interference into account.

More precisely, let us consider a multi-agent system, which consists of two types of agents, type $\B$: BS and type $\D$: devices and an environment $E$. 
Initially, device agents could be either distributed on the edges of a real urban street $E$ (i.e., streets of the city) according to a linear Poisson point process (PPP) or distributed everywhere in $E$ according to a planar PPP. 
Base-station agents can be either placed randomly on $E$ (on streets or everywhere) using a PPP or placed according to given geographic coordinates. 

\subsection{Device agents}\label{rb}
We start by considering a ﬁnite transition system that captures the behavior of a set of $n$ {\em device agents} 
$\D = \{\,d_i \colon i \in \{1,\dots,n\}\,\}$, given by
\begin{equation}
{\rm T}^{d} := \langle {\rm S}^{d}, {\rm Act}^{d}, {\rm R}^{d} \rangle.
\end{equation}
Here, ${\rm S}^{d} :=  \{C_-, C_0,\dots,C_{h_{\rm max}-1}\}\ $ is the {\em state space}, where the state $C_-$ refers to {\em not connected}, $C_0$, refers to {\em connected directly to the BS} (without any hop) and  $C_{h_{\rm max}-1}$ refers to the state {\em connected to BS with $h_{\rm max}$ hops}, where $h_{\rm max}$ is the number of maximum allowed hops. Further, 
$${\rm Act}^{d} := \{ {\rm move, discover, connect_b, connect_d, disconnect} \}$$ 
 denotes the set of {\em actions} that each device agent can perform. Finally, ${\rm R}^{d}$ represents the set of the {\em behavioral rule base} of the agents $\D$.

Formally, each device agent $d_i$ is defined in each time slot by a tuple 
\begin{equation}
\M^{d}_{i,k} := \langle X^{d}_{i,k}, V_{i,k}^{d}, N_{i,k}^{d}, {\rm Act}_{i,k}^{d}, \xi_{i,k}^{d} \rangle.
\end{equation}
Here, $X_{i,k}^{d}$ specifies the agent's {\em location} in terms of coordinates at time $k{\rm d}t$, $V_{i, k}^{d} = v$ represents the agent's moving {\em velocity} and $N_{i,k}^{d}$ represents the {\em knowledge base}, representing what each agent $d_i$ knows about its neighborhood agents and the environment in time slot $k$. $\xi_{i,k}^{d} \in {\rm S}^{d}$ represents the state of agent $d_i$. ${\rm Act}_{i,k}^{d}$ is the set of actions that could be performed by $d_i$ at time $k$.
\subsection{Device-agent behavior}
Agent’s behavior refers to the representation of a process that links the agent’s sensing of its environment to its actions. Let us describe the different behaviors of device agents in terms of mobility, neighbor discovery, connection and disconnection.

\medskip
\subsubsection{Mobility}
Devices move at the same constant speed $v$ repeating indefinitely the following mobility scheme (random-walk model adapted to streets):
\begin{itemize}
    \item A device moves along a road without changing direction until reaching an intersection.
    \item At intersections, the probability $p_d$ to keep the same direction is equal to 0.5 (set to 0 in case of a T crossing intersection) and the probability to take a new direction is $p_l ( p_l = (1-p_d) / (l-1))$, where $l$ is number of lanes throughout the intersection. 
    \item The device moves backward and reverses its direction if a dead-end is reached. 
\end{itemize}

\medskip
\subsubsection{Neighbor discovery}
The discovery process is important to set up connection between agents devices as well as BSs. 
The considered network discovery scheme is \textit{device centric} (without involving BSs):
 \begin{itemize}
    \item Each device agent can discover autonomously and directly neighboring devices within a given communication radius $r_d$.
    \item Each device agent can discover autonomously and directly BS agents within a given communication radius $r_b$.
\end{itemize}
These rules suppose that the emission power of the objects is a constant, and that we do not take into account interference.

\medskip
\subsubsection{Connection}
In order to establish a connection to the network, device agents adhere the following rules:
\begin{itemize}
    \item If a device discovers neighboring BS agents, it will connect to the nearest one.
    \item Else, if it has a neighbor device connected to the network (with less than $h_{\rm max}$ hops), it will connect to it. 
\end{itemize}

\medskip
\subsubsection{Disconnection}
Device agents disconnect from the network according to the following rules:
\begin{itemize}
    \item If the distance between a connected device agent and its associated BS exceeds the communication radius $r_b$.
    \item If the distance between two connected devices (with one hop) becomes higher than the device communication radius $r_d$. 
    \item A device becomes disconnected if its relay device becomes disconnected.
\end{itemize}

\subsection{Rule base}
\label{rb}
\begin{figure}
  \includegraphics[width=\linewidth]{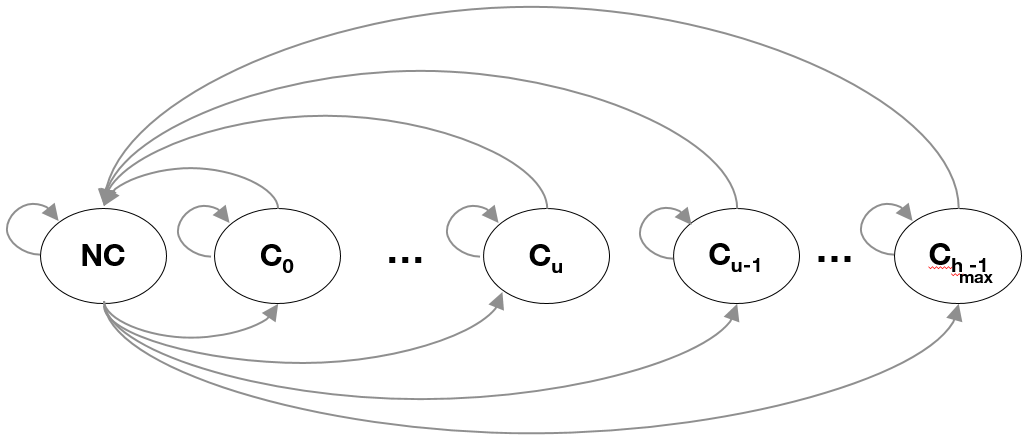}
  \caption{Illustration of device-agent state transitions}
  \label{fig:st}
\end{figure}

The rule base ${\rm R}^d$ 
implements the reactive behavior of device agents as illustrated in Figure~\ref{fig:st}. It allows to select actions to take for agent $d_i$ depending on its current local state $\xi_{i,k}^{d}$ and its knowledge base $N_{i,k}^{d}$.
More specifically, ${\rm R}^d=\{ \Theta\}\ $ where $\Theta( \xi_{i,k}^{d}, N_{i,k}^{d})$ are the {\em active rules} that map the set of states and observations to actions for reactive tasks 
$$\Theta : ( \xi_{i,k}^{d}, N_{i,k}^{d}) \longrightarrow {\rm Act}_{i,k}^{d}.$$
The principal rule-based functions are described as follows. 
\begin{itemize}
    \item Discovery rule: a disconnected agent $d_{i}$ is constantly discovering its environment (the action $\rm discover$ is activated),  
    $$\Theta_{D} : ( \xi_{i,k}^{d}, N_{i,k}^{d})     \longrightarrow {\rm discover}.$$
    \item Connection to BS rule: if a disconnected agent $d_{i}$ discovers a BS agent $b_{j}$ within a Euclidean distance of $r_{b}$ ($\xi_{i,k}^{d} = C_-$, distance($d_{i},b_{j}) < r_{b}$), then agent $d_{i}$ will transit from state $C_-$ to state $C_0$ (the action $\rm connect_b$ will be activated),    $$\Theta_{C^{d}} : ( \xi_{i,k}^{d}, N_{i,k}^{d})     \longrightarrow {\rm connect_b}.$$
    \item Connection to device rule: if a disconnected agent $d_{i}$ discovers a connected agent $d_{j}$ connected to BS via $u$ hops ($u < h_{max}$) and within a Euclidean distance of $r_{d}$ ($\xi_{i,k}^{d} =C_-$, $\xi_{j,k}^{d} = C_u$, distance($d_{i},d_{j}) <  r_{d}$), then agent $d_{i}$ will transit from state $C_-$ to state $C_{u+1}$ (the action ${\rm connect_d}$ will be activated),    
$$\Theta_{C^{b}} : ( \xi_{i,k}^{d}, N_{i,k}^{d})     \longrightarrow {\rm connect_d}.$$
    \item Disconnection from BS rule: if the distance between a connected agent $d_{i}$ and its associated BS agent $b_{j}$ becomes higher than $r_{b}$ ($\xi_{i,k}^{d} = C_0$, distance($d_{i},b_{j}) > r_{b}$), then agent $d_{i}$ will  transit from state $C_0$ to state $C_-$ (the action $\rm disconnect$ will be activated),    $$\Theta_{D^{b}} : (\xi_{i,k}^{d}, N_{i,k}^{d})     \longrightarrow {\rm disconnect}.$$
    \item Disconnection from device rule (1): if the distance between a connected agent $d_{i}$ and its relay device agent $d_{j}$ ( connected to BS via $u$ hops ) becomes higher than $r_{d}$ ($\xi_{i,k}^{d} = C_{u+1}$, $\xi_{j,k}^{d} = C_u$, distance($d_{i},d_{j}) >  r_{d}$ ), then agent $d_{i}$ will transit from  $C_{u+1}$ to $C_-$ (the action $\rm disconnect$ will be activated),    $$\Theta_{D^{d}_1} : ( \xi_{i,k}^{d}, N_{i,k}^{d})     \longrightarrow {\rm disconnect}.$$
     \item Disconnection from device rule (2): if a device agent $d_{i}$ is connected to a relay agent $d_{j}$ that was connected to BS via $u$ hops but looses connection,  then $d_{i}$ will become also disconnected ($\xi_{i,k}^{d} = C_{u+1}$, $\xi_{j,k}^{d} = C_-$), then agent $d_{i}$ will transit from state $C_{u+1}$ to state $C_-$ (the action $\rm disconnect$ will be activated),    $$\Theta_{D^{d}_2} : ( \xi_{i,k}^{d}, N_{i,k}^{d})     \longrightarrow {\rm disconnect}.$$   
\end{itemize}

\subsection{Base station agents}
As above, we consider a ﬁnite transition system that capture the behavior of a set of $m$ {\em BS agents} $\B = \{\,b_j \colon j \in [1,m]\,\}$, i.e., 
\begin{equation}
{\rm T}^{b} = \langle {\rm S}^{b}, {\rm Act}^{b}, {\rm R}^{b}
\rangle.
\end{equation}
Here, the state space is given by ${\rm S}^{b} =  \{\rm on, off\}$, to refer to states {\em on} and {\em off}, respectively. Further, ${\rm Act}^{b} = \{ {\rm cover} \}$ denotes the set of {\em actions} that each BS agent can perform. Actions are limited, in this model, to connectivity coverage for connecting networking devices together as well as connecting the devices to the internet. 
${\rm R}^{b}$ represents the set of the {\em behavioral rule base} of agents $\B$. For the rest of this paper, we assume BS agents to be always in state {\rm On} and are constantly covering the $r_b$ radius surrounding areas. 

\section{Agent-based simulations for D2D}
\label{sec:simu}

In this section we present more details on the implementation of our agent-based simulation and on the considered use cases.
\subsection{Simulation and algorithms}
Let us denote by 
\begin{equation}
\Prms := \{{\rm d}t, r_b, r_d, \lambda_b, \lambda_d, v \}
\end{equation}
the set of key model parameters where ${\rm d}t$ represents the elapsed time in each step, $r_b$ and $r_d$ represent the transmission radius of BS agents and the communication radius of device agents, respectively. $\lambda_b$ is the intensity of BS agents, $\lambda_d$ is the intensity of device agents (agent/$km^2$ for the planar homogeneous PPP of the first use case, or agent/$km$ for the linear homogeneous PPP of the second use case on streets, cf.~Section~\ref{sec:usecases}) and $v$ denotes agents speed ($km/h$). Our simulation is performed over steps and each step corresponds to a time point $k{\rm d}t$. In the following we will denote by $\M_k$ the model at step $k$. 
It represents the environment, the agents and all their attributes (coordinates, states, etc.) at step $k$. 

\begin{table}[h]
\begin{center}
\begin{tabular}{ |p{0.5in}|p{2.5in}| } 
\hline
Parameter & Description \\ 
\hline
$dt$  & Elapsed time in each step (s)\\
\hline
$r_b$ & Transmission radius of BS agents\\
\hline
$r_d$ & Communication radius of device agents\\
\hline
$\lambda_b$ & Intensity of BS agents\\
\hline
$\lambda_d$ & Intensity of device agents\\
\hline
$v$  & Speed of agents ($km/h$) \\
\hline
\end{tabular}
\end{center}
\caption{Simulation parameters}
  \label{tab:param}
\end{table}

In the simulation, we first integrate a real city map and then the agents. After that we run the function Step($\M_k$), that updates the variables of the model, taking it from a step $k$ to the next step $k+1$ for a number $k_{\max}$ of iterations. Algorithm~\ref{algo:main} describes the entry function of the simulation.
\begin{algorithm}[h]
   \SetKwInOut{Input}{Input}
   \SetKwInOut{Output}{Output}
   \caption{Main($\Prms,k_{\max} $): Main function}\label{algo:main}
   \Input{set of parameters $\Prms$ and maximum number of steps $k_{\max}$}
   \Output{The state of a randomly generated model at time $k_{\max}{\rm d}t$}
   $\Map \gets $ GenerateEnv(Lat, Long)\;
   $\B \gets$ GenerateBs($\lambda_b$)\;
   $\D \gets$ GenerateDevices($\lambda_d$)\;
   $\M_0 \gets (\Prms, \Map, \B, \D)$\;
   \For{$k \in \{1, \ldots ,k_{\max}\}$}{
        $\M_k \gets \operatorname{Step}(\M_{k-1})$\;
   }
   \KwRet{$\M_{k_{\max}}$}
\end{algorithm}
The function ${\rm GenerateEnv}$ integrates the real urban environment on which simulations will be performed. The functions ${\rm GenerateBs}$ and ${\rm GenerateDevices}$ return the sets of agents $\B$ and $\D$ using homogeneous PPP with parameters $\lambda_b$ and $\lambda_d$, respectively. \\
The core function of our simulation, called step function, is described in Algorithm~\ref{algo:step}. At each step, device agents move
according to the mobility model described in Section~\ref{rb}. The function ${\rm Update}(d_i)$ checks if agents still connected to their associated BS or device relays after movement and activates the action {\rm disconnect}, as described in Section~\ref{rb}, if needed. If an agent $d_i$ is not connected, then the functions ${\rm Discover}(d_i)$ and ${\rm Connect}(d_i)$ are activated to discover the neighborhood and connect to the network.   

\begin{algorithm}[h!]
   \SetKwInOut{Input}{Input}
   \SetKwInOut{Output}{Output}
   \caption{Step Function}\label{algo:step}
   \Input{The model $\M_{k-1}$ at step $k-1$}
   \Output{The model $\M_{k}$ at step $k$}

   \For{$d_i \in \D$ }{
        $X_{i,k}^d \gets \operatorname{Move}(d_i,V,X_{i,k-1}^d, {\rm d}t)$; \\
        $\operatorname{Update}(d_i)$;\\
        \If{($\xi_{i,k}^{d} = C_-$)}{
         \If{($\operatorname{Discover}(d_i) \neq null$)}
             { $\operatorname{Connect}(d_i)$;
             }
            }
   }
   $\M_{k} \gets (\Prms, \Map, \D, X_{i,k}^d, \xi_{i,k}^d)$\;
   \KwRet{$\M_k$}\;
\end{algorithm}

\subsection{Use cases}
\label{sec:usecases}

For applying the multi-agent simulation to our D2D context, we considered 
two use cases.
First, we will show that agent-based simulation can be used to determine the coverage extension of a BS thanks to multi-hop D2D.

Second, we will show that we can import real street systems and implement mobility of devices in an agent-based simulation to determine the improvement of the connection time of devices to the BS thanks to multi-hop D2D. This will in particular show the importance of the street system morphology.
\begin{figure*}[ht!]
     \begin{center}
        \subfigure[Paris street network]{%
            \label{fig:paris}
            \includegraphics[width=0.5\textwidth]{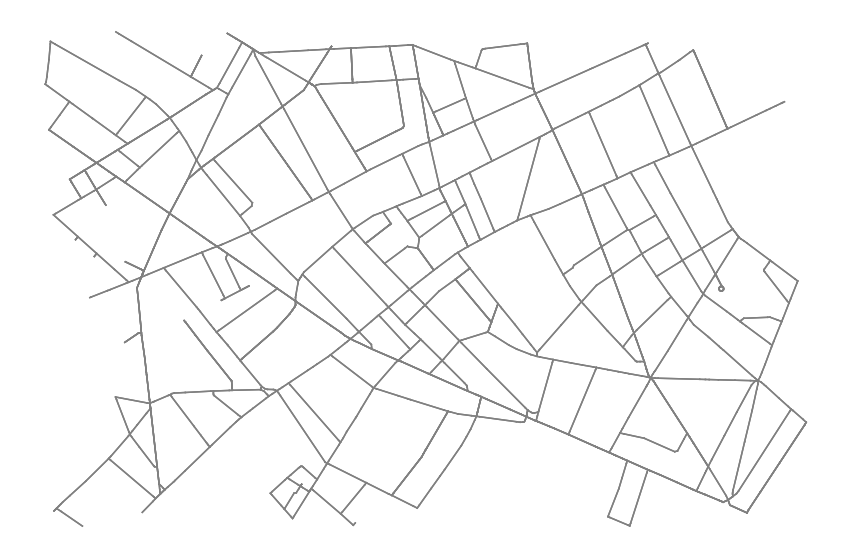}
        }%
        \subfigure[Lyon street network]{%
           \label{fig:lyon}
           \includegraphics[width=0.5\textwidth]{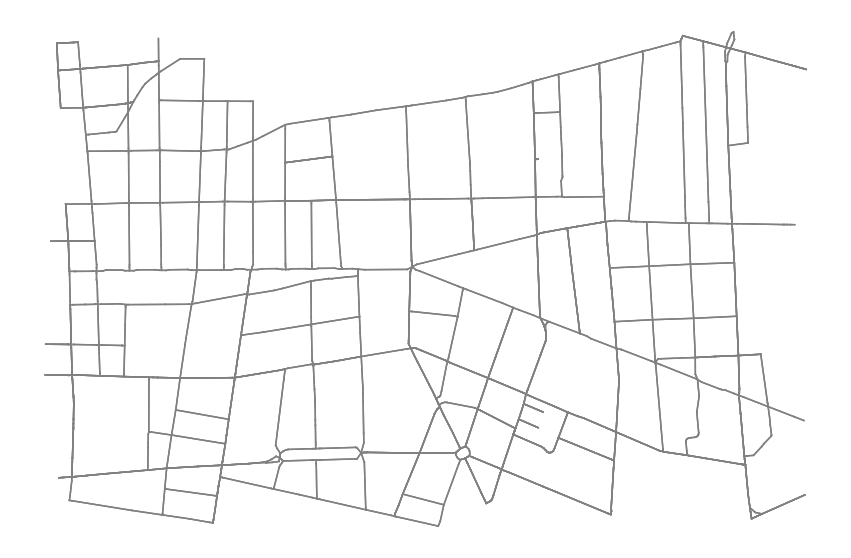}
        }\\ 
        \subfigure[Houston street network]{%
            \label{fig:houston}
            \includegraphics[width=0.5\textwidth]{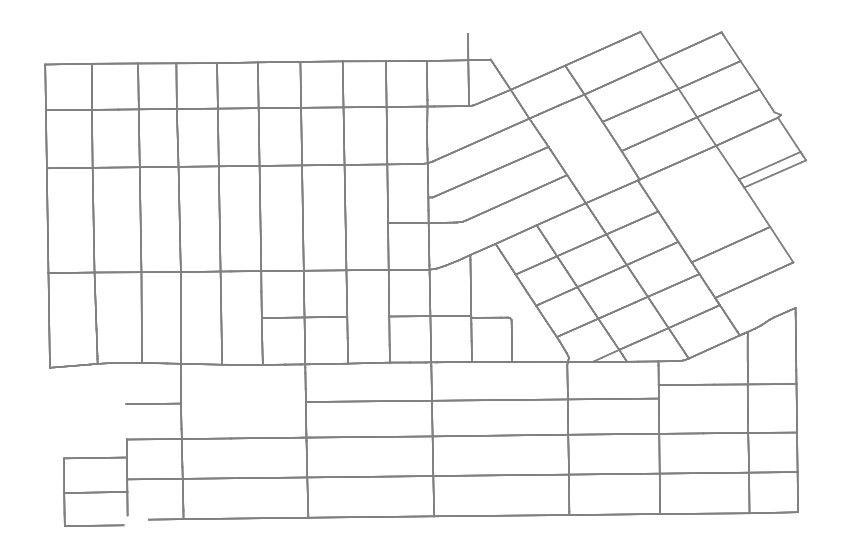}
        }%
        \subfigure[Xian street network]{%
            \label{fig:xian}
            \includegraphics[width=0.5\textwidth]{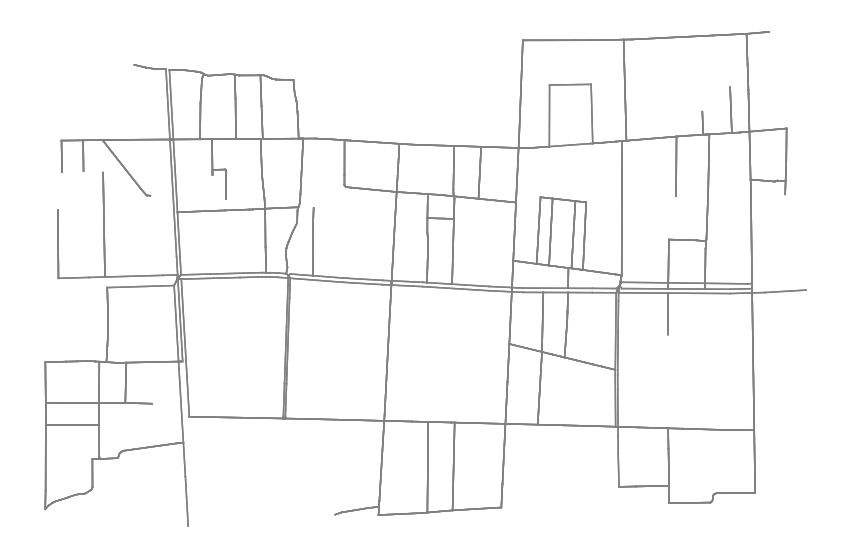}
        }%
    \end{center}
    \caption{%
         Urban Street Networks from~\cite{OSM}
     }%
   \label{fig:subfigures}
\end{figure*}

\subsubsection{Coverage extension}

For the first use case, we only consider one BS at the center of the simulation window, and devices are distributed in the window as a PPP of intensity $\lambda$. The BS and the devices have a communication radius $r_b$ and $r_d$, respectively. We will call  "relaying hop" (or simply "hop") a communication between two devices, excepting the source and participating in relaying a message to the destination. Thus, a direct communication between a BS and a device has zero relaying hops, while a communication from a BS to a device using one intermediary relaying device has one relaying hop. The question we want to answer is the following. With the help of multi-hop D2D with a maximum number of hops $h_{\max}$, what is the expected area $A$ of the surface covered by the BS (meaning that, if a device is inside this surface, it is connected to the BS with a maximum of $h_{\max}$ hops)?
For $h_{\max} = 0$, the surface is a disc of radius $r_d$ centered at the BS, so $A = \pi r_b^2$.
For $h_{\max} = 1$, if $B_{r_d}(x)$ is the ball of radius $r_d$ centered at $x$, and $e_1$ is a unit vector, we have the following theorem.
\begin{theorem}\label{thm_1}
Let a base station of communication radius $r_b$ be placed at the origin, and let devices of communication radius $r_d$ be distributed in $\R^2$ as an {\em homogeneous Poisson point process} $X$ of intensity $\lambda$. Then, the {\em expected} area $A$ covered by the base station augmented by one-hop D2D is given by
$$A = \pi (r_b+r_d)^2 - 2 \pi \int_{r_b}^{r_b+r_d} s \exp(-\lambda |B_{r_b} \cap B_{r_d}(se_1)|) {\rm d} s.$$
\end{theorem}

\begin{proof}
Let $o$ be the origin of $\R^2$. 
We will denote by $x\leftrightarrow y$ the event that $x$ is connected to $y$ with at most one relaying hop. Then, we have that
\begin{align*}
A &= |B_{r_b}| + 
\int_{B_{r_b+r_d}\setminus B_{r_b}} \Prob (x\leftrightarrow o) {\rm d}x\\
&= |B_{r_b}| + \int_{B_{r_b+r_d}\setminus B_{r_b}} (1-\Prob(x\nleftrightarrow o)) {\rm d}x\\
&= |B_{r_b+r_d}| - \int_{B_{r_b+r_d}\setminus B_{r_b}} \Prob(X\cap(B_{r_b}\cap B_{r_d}(x)) = \emptyset) {\rm d}x\\
&= |B_{r_b+r_d}| - \int_{B_{r_b+r_d}\setminus B_{r_b}} \exp(-\lambda |B_{r_b}\cap B_{r_d}(x)|) {\rm d}x,
\end{align*}
and hence,
\begin{align*}
A&=|B_{r_b+r_d}| - 2\pi \int_{r_b}^{r_b+r_d} s \exp(-\lambda |B_{r_b}\cap B_{r_d}(se_1)|){\rm d} s,
\end{align*}
which concludes the proof.
\end{proof}
In Section~\ref{sec:results} we present simulation results for $2 \leq h_{\max} \leq 7$.

\medskip
\subsubsection{Implementing mobility on various real street systems}
We use four different cities, presenting street systems of various morphology: Paris and Lyon have old European-style street systems, Houston has a Manhattan-style street system, while Xian has a nested Manhattan-style street system, as depicted in Figures~\ref{fig:paris},~\ref{fig:lyon},~\ref{fig:houston}, and~\ref{fig:xian}, respectively. 

We positioned one BS in the simulation window in order to cover approximately the same length of streets.
We then measured the mean time of connection of the devices to the BS through multi-hop D2D, with $0 \leq h_{\max} \leq 5$.

\section{Simulation study}
\label{sec:results}
\subsection{Simulation environment and settings}
Our proposed model is based on the {\em NetLogo} framework~\cite{netlogo}, which is a multi-agent programmable modeling environment for simulating complex systems that we have extended to integrate and visualize real street systems. 
For all simulations, unless otherwise stated, parameters are set as follows: 
total streets length = $35 km$ and total simulation time = $600s$, $v= 5 km/h$ (average pedestrian speed), $r_b = 300m$ (a plausible range for future access points), $r_d = 100m$.

\subsection{Coverage extensions}
\begin{figure}
  \includegraphics[width=\linewidth]{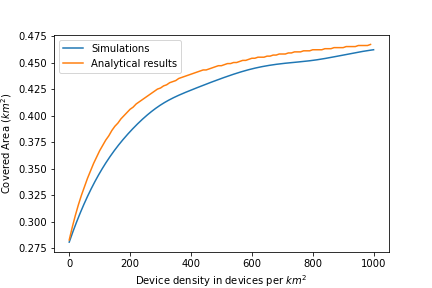}
  \caption{Comparison between simulation and analytical results (Theorem~\ref{thm_1}) for the covered area ($h_{\max}=1$).}
  \label{fig:ca_theo_simu}
\end{figure}
\begin{figure}
  \includegraphics[scale=0.6]{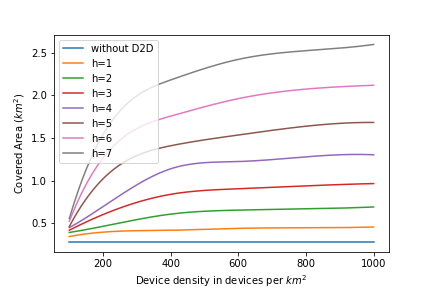}
  \caption{Simulation results for the covered area  for $h_{\max} \leq 7$.}
  \label{fig:coverage-ex}
\end{figure}

As depicted in Figure~\ref{fig:ca_theo_simu}, we could quite faithfully reproduce the curve for the case $h_{\max} = 1$. Let us note that the underestimation of the simulation curve compared to the analytical results from Theorem~\ref{thm_1}  can be, to a large extend, attributed to the spatial discretization scheme used within NetLogo. The simulation curves for $2 \leq h_{\max} \leq 7$ are presented in Figure~\ref{fig:coverage-ex}. Here we observe a sharp increase in coverage with respect to the number of allowed hops for small device intensities, followed by a saturation effect for larger device intensities. 

\subsection{Mean connection time to base station}
\begin{figure*}[ht!]
     \begin{center}
        \subfigure[Mean connection time to the BS in Paris]{%
            \label{fig:ct_paris}
            \includegraphics[width=0.5\textwidth]{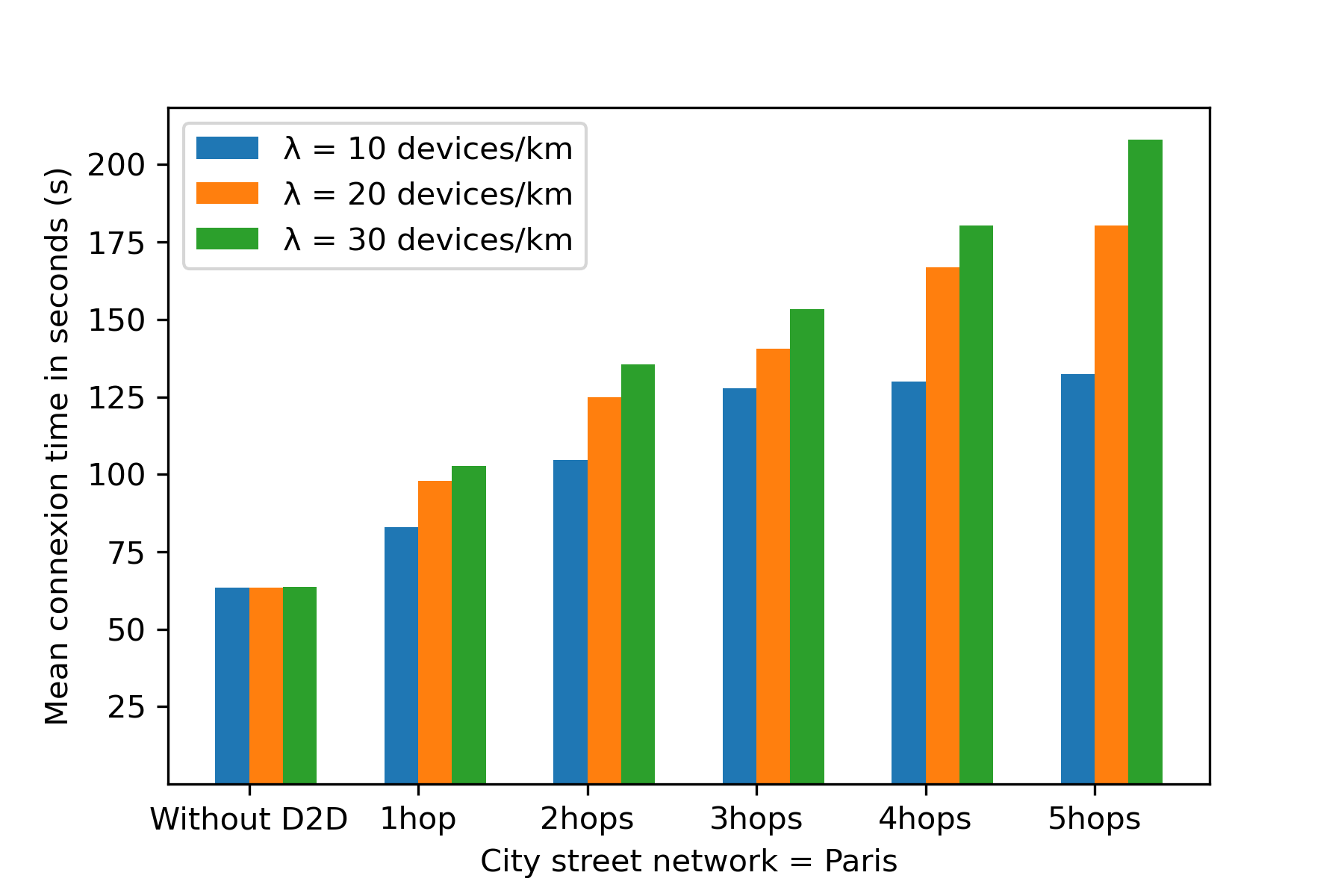}
        }%
        \subfigure[Mean connection time to the BS in Lyon]{%
           \label{fig:ct_lyon}
           \includegraphics[width=0.5\textwidth]{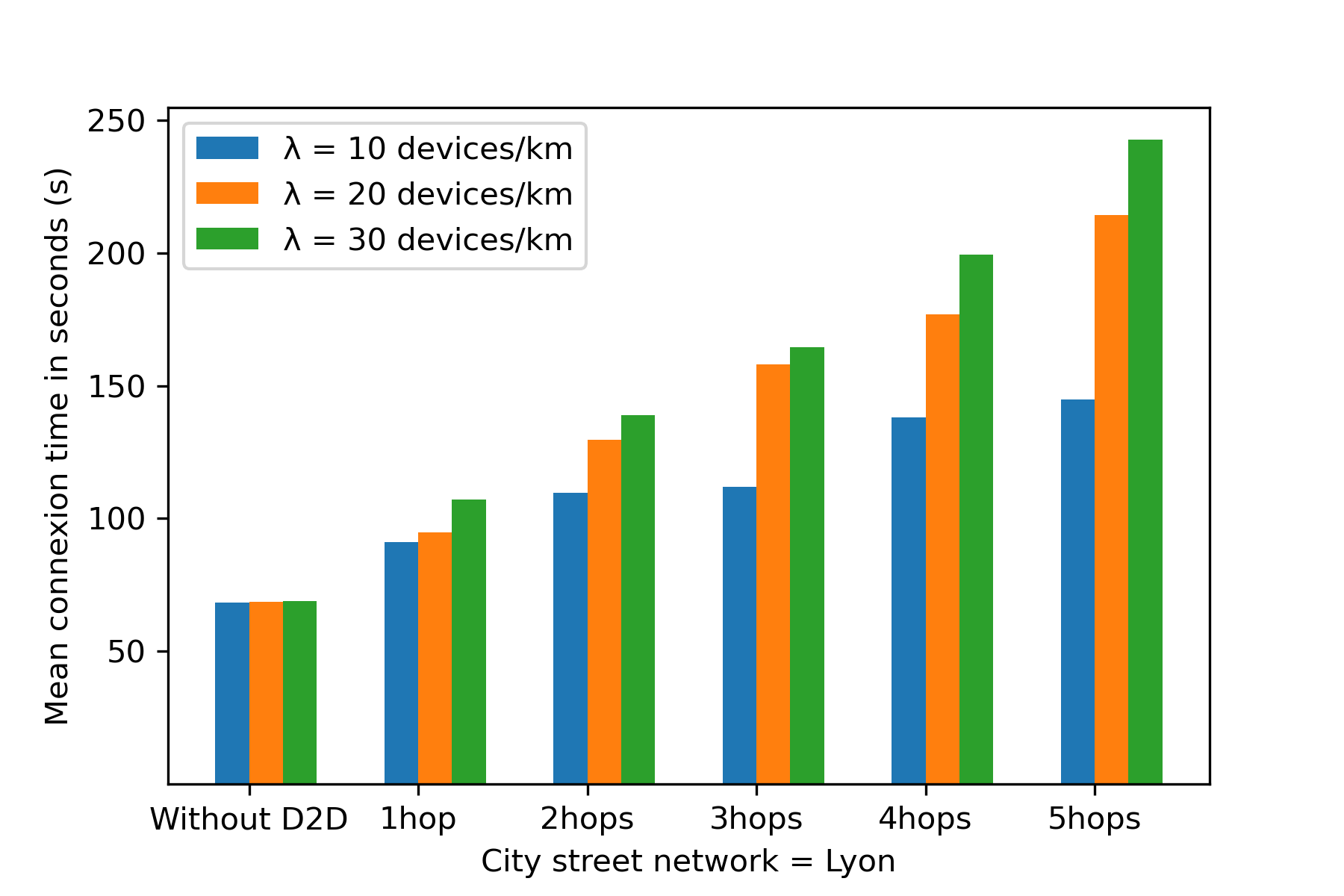}
        }\\ 
        \subfigure[Mean connection time to the BS in Houston]{%
            \label{fig:ct_houston}
            \includegraphics[width=0.5\textwidth]{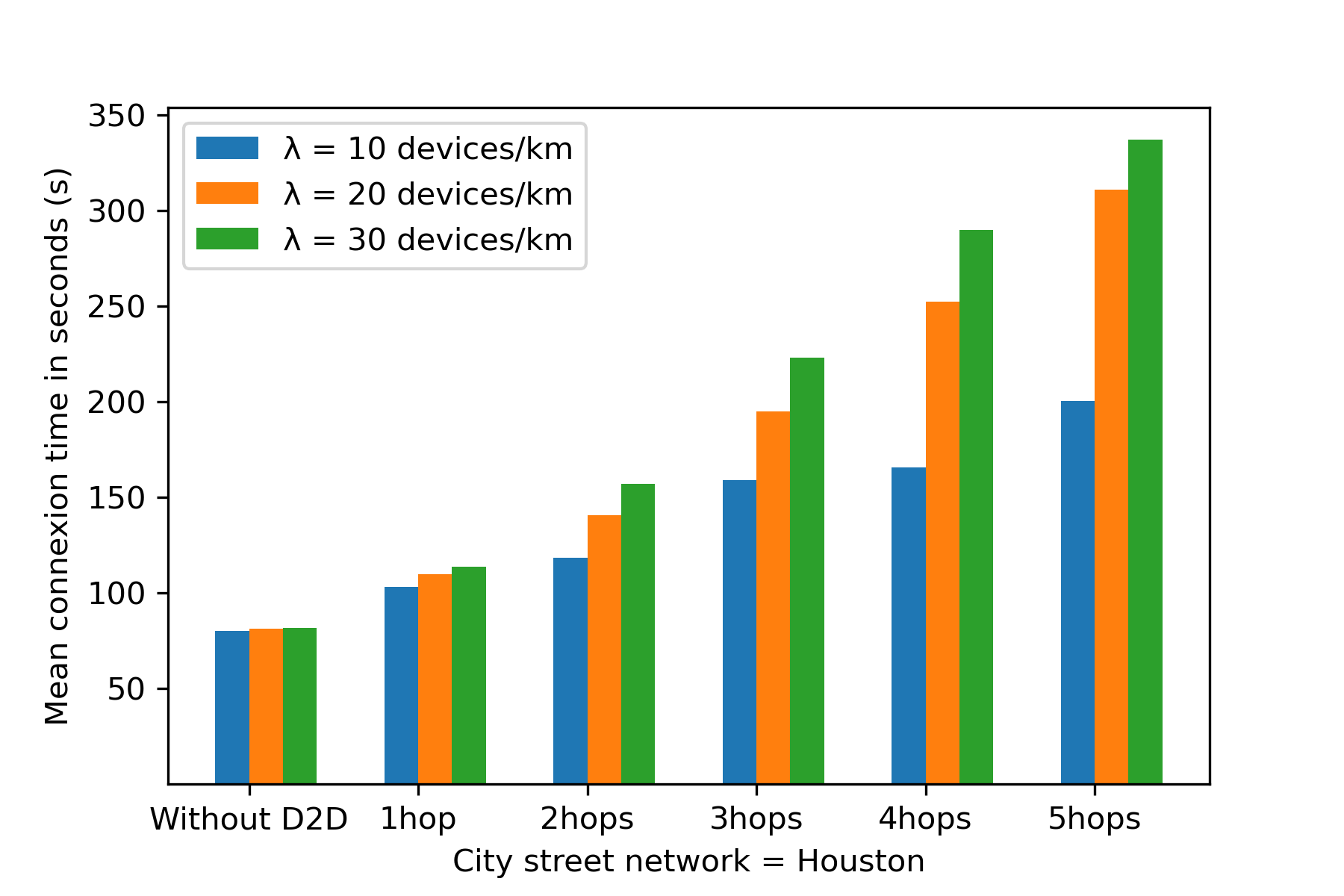}
        }%
        \subfigure[Mean connection time to the BS in Xian]{%
            \label{fig:ct_xian}
            \includegraphics[width=0.5\textwidth]{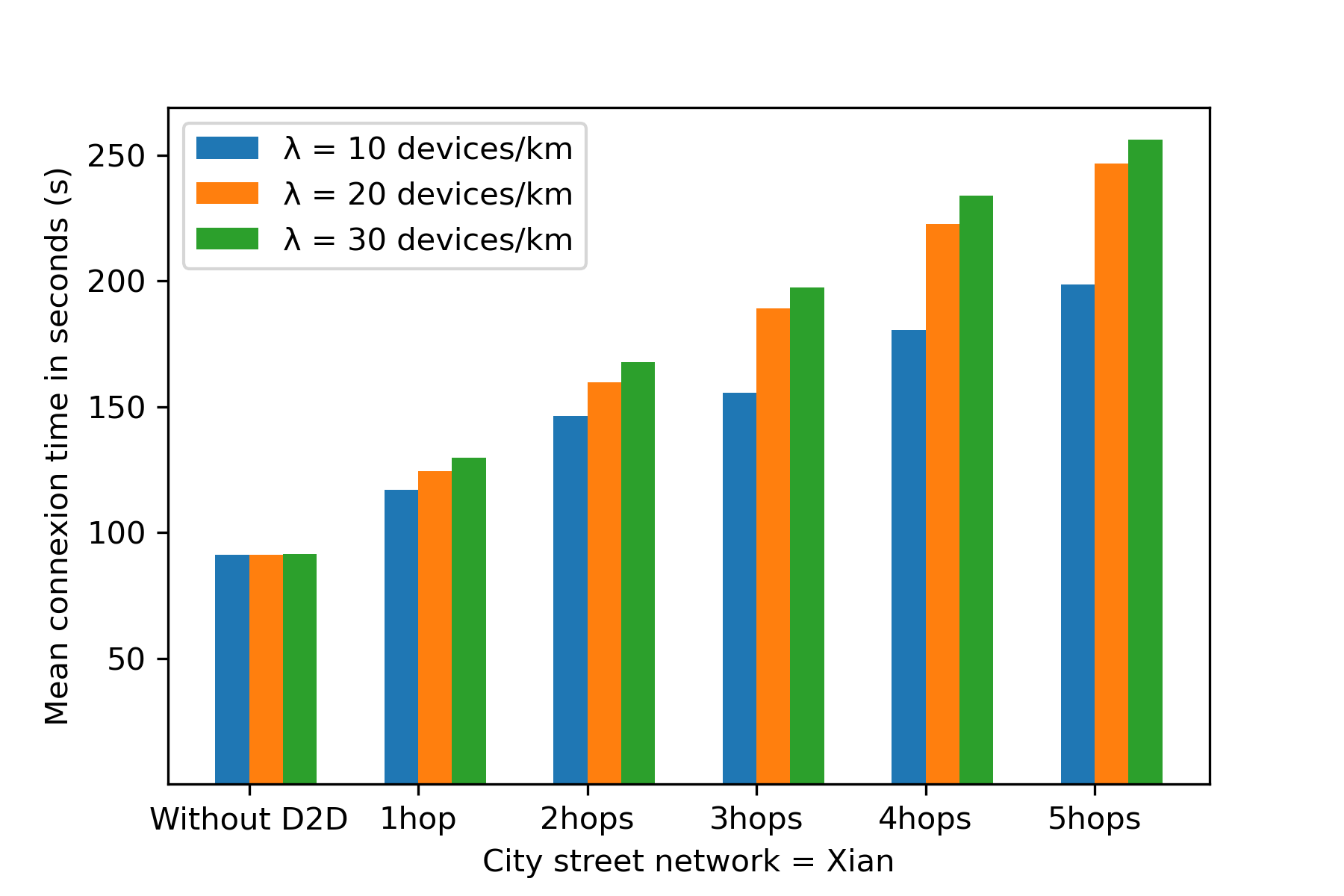}
        }%
    \end{center}
    \caption{%
        Simulation results for the mean connection time to BS.
     }%
   \label{fig:subfigures}
\end{figure*}

In all the figures, we observe that without D2D, the mean connection time to the BS is independent of the devices density and roughly equal to the ratio of the length of street covered by the BS to the total length of streets inside the window. Indeed, the initial placement of the devices follows a PPP, and the random-walk laws do not much alter the uniformity of the distribution of the devices for the time of the simulation.

Then, allowing multi-hop and increasing the density of devices enhances the coverage of the BS, but varies according to the street-system morphology.
One general feature that we can observe on all figures is that, increasing the density of devices from $10$ to $20$ devices/$km$, is in any case more effective than increasing it from $20$ to $30$ devices/$km$.

Figures~\ref{fig:ct_paris} and~\ref{fig:ct_lyon}, featuring European cities, are very similar: for a device intensity of 10 devices/$km$, we have a maximum of mean connection time around $130$ seconds for $5$ hops, and the connection time increases for low device densities and then reaches a plateau for high device densities. The increase of the connection time for $5$ hops from $10$ to $20$ devices/$km$ is quite significant: more than $50$ seconds.

For Houston (cf.~Figure~\ref{fig:ct_houston}), the situation is clearly different: the mean connection time for $5$ hops is almost twice as large as for the two previous cases, and the increasing of the connection time gets larger when increasing the number of hops.

For Xian (cf.~Figure~\ref{fig:ct_xian}), the situation is in between: for instance, while the global profile of the curves look more like the European cities, the mean connection time for $10$ devices/$km$ and $5$ hops is almost as high as in the case of Houston.

\section{Discussion and Conclusion}
\label{sec:ccl}
This paper investigates the application of ABM and simulation for D2D augmented networks. We propose an ABM approach for coverage extensions in D2D with multi-hops and investigate a variety of use cases. First, we show that our approach can predict coverage extension for multi-hop D2D. We prove rigorously the effectiveness of our model and benchmark it with a basic one-hop scenario that is easier to analyze mathematically. Then, we show, particularly, the importance of urban street systems and device mobility to determine the improvement of the connection time of devices to the BS thanks to multiple hops.

This work is an initial effort to apply ABM to D2D networks. It aims to open horizons for more sophisticated use cases, featuring for instance energy consumption, interference, multipath propagation, or multiple BSs.

Refining the mobility aspects could be achieved through more sophisticated mobility tools such as the MATSim simulator~\cite{MATSim}, which includes urban public traffic, or Sumo~\cite{Sumo}, which simulates urban traffic excluding pedestrian, or The ONE~\cite{Theone} as an urban mobility simulation framework. 

As a future work, we aim to conduct more simulations to study other important aspects like D2D connection stability while also considering realistic mobility patterns.

\section*{Acknowledgement}
CH acknowledges the financial support of the
CogniGron research center and the Ubbo Emmius Funds (University of Groningen).

BJ acknowledges the financial support of the German Research Foundation under Germany’s Excellence Strategy MATH+: The Berlin Mathematics Research Center, EXC-2046/1 project ID: 390685689 and the German Leibniz Association via the Leibniz Competition 2020.

\bibliographystyle{IEEEtran}
\bibliography{sigproc}

\end{document}